# Diffuse neutron scattering in relaxor ferroelectric $PbMg_{1/3}Nb_{2/3}O_3$

**Sergey Vakhrushev[a], Alexandre Ivanov*[b], Jiri Kulda[b]**

[a] *Ioffe Physico-Technical Institute, 26, Politekhnicheskaya, 194021 St.-Petersburg, Russia. Fax: +7-812-514 1815; Tel: +7-812-515-6747; E-mail: s.vakhrushev@mail.ioffe.ru*
[b] *Institut Laue-Langevin, 6, rue Jules Horowitz, B.P.156X, 38042 Grenoble, France. Fax: +33-4-76-20-76-88; Tel: +33-4-76-20-74-74; E-mail: aivanov@ill.fr*

*This submission was created using the RSC Article Template (DO NOT DELETE THIS TEXT)*
*(LINE INCLUDED FOR SPACING ONLY - DO NOT DELETE THIS TEXT)*

High energy resolution neutron spin-echo spectroscopy has been used to measure intrinsic width of diffuse scattering discovered earlier in relaxor ferroelectric crystals. The anisotropic and transverse components of the scattering have been observed in different Brillouin zones. Both components are found to be elastic within experimental accuracy of 1 μeV. Possible physical origin of the static-like behavior is discussed for each diffuse scattering contribution.

## Introduction

Relaxor Ferroelectrics (RF) generally mean complex perovskites with the $A(B'B'')O_3$-type cubic unit cell with random occupation of the cation sites A and/or B by unlike-valence cations. These materials were discovered fifty years ago and first named "ferroelectrics with diffuse phase transition" (for a review of early results see [1]). The RF are characterized by dielectric constant of extremely high value associated with significant frequency dispersion. The variation of the diffuse maximum of dielectric constant in broad frequency and temperature ranges gives evidence for an important role of slow (as compared to typical phonon times) relaxation processes in these compounds. To emphasize this fact Cross [2] has baptized such materials Relaxor Ferroelectrics. As a rule, the relaxor behavior is not related to a structural phase transition.

Lead magnoniobate $PbMg_{1/3}Nb_{2/3}O_3$ (PMN) can be considered as a generic relaxor and is the most studied one. At the freezing temperature $T_g$ of about 230 K it exhibits a phase transition into a dipole-glass phase as evidenced by divergence of nonlinear susceptibility [3] and appearance of history dependent effects [4]. There are several theoretical approaches used to describe the properties of relaxors and PMN in particular. Practically all of them are based on the assumption of the existence of slowly relaxing polar nanoregions (PNR) at temperatures much above $T_g$. It was proposed by Burns that these PNR are formed at some temperature $T_d$ (now usually called Burns temperature) much higher than $T_g$ [5], for instance, in the case of PMN $T_d \sim 630K$. Recently it was demonstrated that the formation of the PNR is related to the transverse optic (TO) mode softening [6-9]. The most striking difference of the PNR in relaxors from the precursor clusters in majority of compounds undergoing a structural phase transition is the essential difference of the eigenvectors of the soft mode and PNR. However no exact understanding of the microscopic mechanism of the PNR formation exists at present although several approaches have been proposed for understanding their origin, including dynamical phase transition similar to that in glasses [10], Griffith phase formation [11] and others.

A key knowledge for choosing correct physical model or developing a new one can be gained from the dynamical properties of the PNR studied by various experimental techniques. It was found by ultra-broad-band dielectric spectroscopy that significant frequency dispersion develops in PMN below $T_d$ with characteristic frequency ranging over 13 orders of magnitude - from ~10 GHz [12] close to $T_d$ to less then 1 mHz(!) close to $T_g$. However neither dielectric nor optic spectroscopy allow access to the information on *q*-dependence of the relaxation time.

Neutron spectroscopy is the most direct method to study the microscopic origin of atomic dynamics. A three-axis neutron spectrometer has been used to demonstrate that cooling through the Burns temperature results in crossover of critical dynamics in PMN from soft mode to central peak (CP) regime [6]. The monotonic increase in CP intensity below $T_d$ appeared to be similar to that of dielectric constant suggesting the same dynamic origin for both. With the characteristic frequency range of the dielectric behavior well below typical or even the best three-axis neutron spectrometer resolution that was not at all surprising that the CP was found to be resolution limited with the estimated intrinsic width <0.5meV. This CP-related critical scattering, being oriented perpendicularly to the scattered wave vector *Q*, is known as the "transverse" component in distinction from another so called anisotropic diffuse scattering.

The anisotropic component has been observed in many experiments [13-17]. It appears upon cooling at temperatures well below $T_d$ and is not supposed to be related to dynamics of the polar nano-regions. In the vicinity of the (h00) reciprocal lattice points this diffuse component is seen as a butterfly-shaped intensity profile with the "wings" elongated in the [110]-type reciprocal lattice directions. In recent high-resolution three-axis spectrometer measurements [13] and in a back-scattering experiment [14] this anisotropic scattering was concluded to be purely elastic. A meV-range broad quasi-elastic scattering recently reported in [15] has orders of magnitude lower intensity than the diffuse signal under discussion. An attempt to interpret the anisotropic contribution in terms of scattering on the soft mode propagating in the [1 1 0] direction and polarized in the [1 –1 0] direction [16] appeared to be inconsistent with the lattice symmetry (see *e.g.* [17]) while the proposed very special dispersion relation of the transverse acoustic phonons [18] did not find experimental confirmation.

No adequate explanation has been so far given to the physical nature of neither transverse nor anisotropic diffuse scattering component. In the present work we investigate both of them paying special attention to the search for the intrinsic width of the PNR-related transverse contribution. The expected dynamical response can be probed only by experimental techniques which combine adequate sensitivity both to selected zone in the reciprocal space and excitation energy. The most suitable technical solution is provided by a combination of three-axis (TAS) with spin-echo (NSE) neutron spectrometers successfully developing over the last decade. This hybrid spectrometry allows one to access extremely high resolution in energy, $\Delta E/E \sim 10^{-4} - 10^{-5}$ or about 3 orders of magnitude better than typical TAS resolution, fully maintaining the advantages given by typical TAS momentum resolution, $\Delta Q/Q \sim 10^{-2}$.

**Experimental technique**

High resolution inelastic neutron scattering measurements have been carried out using the precession field neutron spin-echo technique implemented on the IN20 tree-axis spectrometer at ILL, Grenoble (TASSE option with optimal-field-shape superconducting precession coils, for details see *e.g.* [19]). This technique demands polarized neutron beams and consists in measuring the time-of-flight of each individual neutron across the spectrometer by Larmor precession in the special zones with well-defined magnetic field in the spectrometer "arms" before and after the sample. Any change in neutron velocity due to energy transfer in the scattering process is revealed by beam depolarization and is measured by the precession phase difference between the two spectrometer arms.

This "individual neutron" phase difference $\Delta\phi = \omega\tau_F$ is related to the energy transfer $\Delta E = \hbar\omega$ through the Fourier time parameter $\tau_F$:

$$\tau_F = (2\gamma m_n^2/h^2)\cdot(\Phi_f/k_f^3) = (2\gamma m_n^2/h^2)\cdot(\Phi_i/k_i^3) \quad (1)$$

where the gyromagnetic ratio $\gamma = 2\mu_n/h = 1.83246\cdot10^7$ T$^{-1}\cdot$s$^{-1}$ includes the neutron magnetic dipole moment $\mu_n$, the neutron velocity $v_n = hk/m_n$ and the precession field integral $\Phi = \int B\cdot dl$, equal just to $BL$ for a homogeneous field of length $L$. The indices $i$ and $f$ refer to the incident and scattered (final) neutron states, respectively.

The experimental procedure consists in measuring polarization of the scattered beam as function of small variations of magnetic field (coil current) in the vicinities of several field values defining the Fourier time (1). The measured intensity will be then presented by sinusoidal-shape curves with a period corresponding to one full turn of neutron spin in the precession coil. The energy width of a dispersionless signal will spread out the initial beam polarization and thus reduce the amplitude of the sinusoid or the contrast. In the case of diffuse or quasi-elastic scattering, when the energy profile of the signal at any momentum transfer $Q$ is expected to be a Lorentzian with FWHM = $\Gamma$, the measured polarization contrast, defined as the sinusoid amplitude divided by its average, should ideally decrease as $P \sim \exp(-\Gamma\tau_F/2)$.

A pyramid-shaped lightly brown-colored transparent single crystal PMN of about 8 cm$^3$ in volume has been used in the reported measurements. It had some optically inhomogeneous regions at the bottom of the pyramid. The crystal was wrapped in a shaped piece of Nb-foil and mounted inside cylindrical heating element of a vacuum furnace with its [001] crystallographic axis perpendicular to the horizontal scattering plane.

Measurements have been performed at two different neutron wave vectors $k_I = 2.662$ Å$^{-1}$ and $k_I = 4.1$ Å$^{-1}$ allowing for effective filtering of high-order scattering harmonics in monochromatic beam. The two experimental settings are characterized by different three-axis resolution and accessible range of Fourier times. Each selected current in the precession coils (or Fourier time value) at each spectrometer configuration has been validated by fine tuning of spin flippers and a number of correction coils in order to optimize the spin-echo contrast on an undoubtedly elastic scattering like Bragg diffraction. Given the optimization procedure is accomplished, the measurements of neutron spin-echo on the scattering from sample have been repeated at different sample temperatures with the three-axis spectrometer set to selected points in the sample reciprocal space. The resulted spin-echo contrast is to be corrected for the instrumental effects derived from the instrument alignment and tuning.

We have noticed certain difference in the instrumental tuning performed, from one side, on a Bragg reflection and, from the other side, on diffuse scattering itself (at room temperature when it is strong enough for the alignment procedure and was expected to be almost perfectly elastic) or powder-like diffraction contribution observed near the (010) Bragg reflection (see below). This is probably due to the fact that angular divergent diffuse or powder scattering was much less sensitive to the details of the spectrometer tuning than a narrow Bragg reflection.

From the ensemble of the alignment runs we concluded that the used experimental set-up did not allow reliable determination of the quasi-elastic width below 1 μeV what should be considered as the level of sensitivity or instrumental resolution.

Special attention has been paid to a possible influence of the electric current in the oven heating element, which could attain values as high as 50 A, on neutron beam depolarization. We compared the spin-echo signal measured without sample, on the direct beam, with and without current in the heating element. The results obtained with the coil current of 42 A at $k_I = 2.662$ Å$^{-1}$ are shown in Fig.1. Although the presence of the heating current results in the overall decrease of the polarization ratio in the whole range of the measured decay times, its effect is reduced at higher $\tau_F$, so that the corrected curve produces a "negative width" of ~0.3 μeV. This value, corresponding to the almost maximal used current, appears to be below the estimated instrumental resolution.

**Results and discussion**

The principal goal of the measurements with $k_I = 2.662$ Å$^{-1}$ was to exploit good three-axis spectrometer resolution in momentum space at this relatively small neutron wave vector in order to facilitate separation of the two diffuse scattering components - transverse and anisotropic. Moreover, as it follows from (1), the accessible Fourier-time range can be made broader as compared to higher neutron energies what may be favorable for the instrument sensitivity.

On the other hand, this relatively short wave vector does not allow one to attain Bragg point (030) where the transverse contribution from polar nano-regions has been observed [6]. That is why we concentrated on the Brillouin zone around Bragg point (010) where the transverse signal is expected to be visible as well. The point (020) has different type of the structure factor and correspondingly much higher Bragg intensity masking diffuse scattering close to this point.

Two-dimensional intensity maps have been measured in the usual three-axis configuration, without spin-echo analysis of neutron polarization, in the vicinity of the different Bragg reflections. As shown in Fig.2a, around the point (010) at room temperature two distinct signals can be distinguished: a butterfly-like with "wings" elongated in the directions close to [1 1 0]-type and a transverse-like component along [1 0 0], the latter on a first glance resembling the PNR-related intensity [6]. On heating to 573 K the "butterfly" has disappeared, as it was expected for the anisotropic diffuse scattering studied as function of temperature in Ref.[20]. At the same time the intensity "stripe" perpendicular to the reciprocal lattice vector [010] has remained practically unchanged, Fig.2b, while it was expected to be strongly suppressed at this temperature slightly below $T_d$. More detailed inspection of the scattering intensity reveals that this "transverse" contribution does not follow a straight line but rather concentrates around the circle centered on the reciprocal space origin, as powder contribution would do. Similar almost temperature independent scattering concentrating on the circle arc perpendicular to $Q$ has been found near the Bragg point (120). We concluded that this practically temperature-independent scattering originates from assembly of small crystallites oriented in a texture-like manner around the main crystal grain. Then the temperature-dependent transverse component considered earlier in [6] appears to be masked by the temperature-independent contribution related to the crystal imperfections. This conclusion was corroborated by additional measurements with a neutron-absorbing mask hiding a part of the optically inhomogeneous regions in the crystal.

The measurements in the vicinity of the reciprocal lattice point (030) have been performed with the neutron wave vector $k_I = 4.1$ Å$^{-1}$. The temperature dependent transverse diffuse scattering has been recovered, as shown in the Fig.3a, mainly due to its order of magnitude higher intensity in this Brillouin zone. At the same time the Bragg contribution is reduced so that the temperature independent contamination due to crystal imperfections appears to be much lower. In agreement with earlier observations this central peak due to PNR-related scattering survived up to Burns temperature while the anisotropic component was already unobservable at 473 K, that is ~150 K below $T_d$, as illustrated by the Fig.3b.

In the following we discuss separately the results obtained in the two Brillouin zones with respect to the two distinct, anisotropic and transverse, components of the diffuse scattering.

**(010) Brillouin zone. Anisotropic component.**

Measurements of neutron spin-echo signal with the three-axis spectrometer set to the "transverse" scattering point $Q = (0.05\ 1\ 0)$ at two essentially different temperatures 300 K and 573 K, Fig.4a, have demonstrated no inelastic contribution, as it was expected for purely elastic diffraction scattering on the crystal texture. Similar measurements on a butterfly "wing" at $Q = (0.04\ 1.05\ 0)$, Fig.4b, result in the same conclusion for the anisotropic diffuse scattering too within the sensitivity of the present experiment. For completeness we note that a test measurements performed at room temperature in the other zone (030) at the scattering wave vector $Q = (0.06\ 3.05\ 0)$ gave the same result - the anisotropic scattering appears to be elastic. This is not surprising because, as it was mentioned above, the anisotropic component was claimed to be elastic in earlier neutron scattering experiments [13,14]. However, no microscopic origin has been proposed for this scattering.

In order to elucidate its possible physical nature, we suggest that local ferroelectric polarization in relaxors is accompanied by local microscopic deformations. Then these inhomogeneous elastic deformations should result in the Huang scattering. We have used the standard theory [21] to describe the wave vector dependence of the scattering intensity as:

$$I_H(\mathbf{Q}) = N_d\ |F(\tau)|^2\ \frac{Q^2}{q^2} \sum_{\alpha=1}^{\nu} (m_i C^{-1}_{\mathbf{q}ij} P_{\alpha jl} n_l)^2$$

where $N_d$ is the number of defects, $F(\tau)$ is the structure factor of the Bragg reflection at the reciprocal lattice vector $\tau$, $\alpha$ - defect orientation index, $\nu$ - total number of possible defect orientations, $Q$ is scattering vector (unit vector $m = Q/|Q|$), $q = Q - \tau$ is the reduced wave vector (unit vector $n = q/|q|$) and the tensor $C_{qij} = c_{ikjm} n_k n_m$ is derived from the assembly of the elastic constants $c_{ikjm}$.

The only fitting parameter here is the elastic dipole force tensor $P$. We have constructed $P_{\alpha jl}$ for different possible symmetries (tetragonal, orthorhombic and rhombohedral) and found that the best agreement with the experimental data can be achieved in the tetragonal case with the ratio of 2 : -1 for the tensor components:

$$P_{1ij} = \begin{pmatrix} 2p & 0 & 0 \\ 0 & -p & 0 \\ 0 & 0 & -p \end{pmatrix};\ P_{2ij} = \begin{pmatrix} -p & 0 & 0 \\ 0 & 2p & 0 \\ 0 & 0 & -p \end{pmatrix};\ P_{3ij} = \begin{pmatrix} -p & 0 & 0 \\ 0 & -p & 0 \\ 0 & 0 & 2p \end{pmatrix}$$

The experimental and calculated intensity maps are presented in Fig.5 demonstrating good agreement of the butterfly-like scattering patterns. This gives evidence for Huang scattering origin of the anisotropic diffuse component. Huang scattering is produced by *static* micro-deformations in the crystal and it is proved to be elastic.

The results of a more detailed study of the anisotropic diffuse scattering will be published elsewhere.

**(030) Brillouin zone. Transverse component.**

The transverse diffuse scattering has been measured with spin-echo set-up in a broad range between room temperature and 573 K, including scans at several intermediate temperatures. The result for the scattering vector $Q = (0.05\ 3\ 0)$ are presented in Fig.6. It should be concluded that this scattering component does not show inelastic contamination within the experimental resolution of ~1 μeV, marked by lines in the figure, similarly to the anisotropic diffuse scattering.

This conclusion seems to be in contradiction with the dielectric spectroscopy data demonstrating frequency response up to GHz (or several μeV) range in approaching $T_d$. One of the reasons may be related to the mentioned crystal imperfections giving rise to diffraction contribution in the transverse signal. At high temperatures the "true" transverse response decreases and close to $T_d$, where its widths should increase, the measured intensity may be overwhelmed by the diffraction scattering. This could be verified with a better quality crystal.

However, the obtained results suggest searching for a possible physical origin of the observed effect. One of the probable explanations is related to the fact that the eigenvectors of the ionic displacements associated with the central peak (transverse) scattering can be decomposed into two different terms with only one of these terms being related to the polarization fluctuations. Indeed, in the paper [22] measurements of the structure factor of the CP were used to determine the eigenvectors of the corresponding ionic displacements. It was demonstrated, that the eigenvectors are neither purely optical nor acoustic. Later Hirota *et al.* [23] have proposed to separate these displacements onto an optic-like component resulting from freezing of the soft mode and the second component called by the authors the phase-shifting mode. Eigenvectors of the second component correspond to static lattice deformation.

Only displacements of the optic type, center of masses conserving, will directly contribute to the dielectric response. However in the scattering experiments both components are observable. At room temperature in the vicinity of the reciprocal lattice point (030) the two displacements are expected to give nearly the same relative intensities, while the temperature dependence of their ratio is not known. In the case if the phase-shifting contribution to the ionic displacements prevails at high temperatures, the faster relaxation near Burns point, clearly indicated by the dielectric spectroscopy, will not be observable in our experiment.

The pure elastic origin of the discussed diffuse scattering can explain a puzzling scaling relation observed in an earlier experiment [4]. In the case of scattering on the fluctuations of thermodynamic order parameter, the dynamic structure factor $S(\mathbf{q},\omega)$ is related to the dynamic susceptibility $\chi(\mathbf{q},\omega)$ as:

$$S(\mathbf{q},\omega) = (h/\pi)(n(\omega)+1)\cdot\chi(\mathbf{q},\omega)$$

where $n(\omega)$ is the Bose factor [17]. At high temperatures ($kT \gg \hbar\omega$) $n(\omega) \sim T/\omega$ and the energy integrated scattering intensity (as measured in X-ray experiment, or in the case of neutron measurements of the CP, when the energy resolution is much broader than the intrinsic CP width) at $q = 0$ should be proportional to temperature: $I(q=0) \sim T\chi^0$, where $\chi^0$ is the static susceptibility. In the mean field approximation the product $\chi^0\cdot\kappa^2 \sim I(q=0)\cdot\kappa^2/T$, where $\kappa^2$ is the square of the inverse correlation length, is not expected to vary with temperature. Instead, in the paper [4] the other product $I(q=0)\cdot\kappa^2$ was found to be temperature independent in the whole region $T_g < T < T_d$ as if the factor $1/T$ originating from the Bose-factor for low-frequency excitations was missing in the experimental scaling relation. This factor is indeed not present in the case of scattering on a static pattern.

**Conclusion**

We have investigated diffuse scattering in a relaxor ferroelectric using high energy resolution neutron spin-echo technique. Both transverse and anisotropic components of the scattering have been found to be elastic within experimental accuracy of 1 μeV. The result for the anisotropic butterfly-like component was anticipated by the previous research. We have suggested its physical origin in Huang scattering on static micro-deformations of the crystal lattice. The transverse scattering related earlier to the dynamic fluctuations of the ferroelectric polarization in the polar nano-regions seems to be in contradiction with the results of the dielectric spectroscopy. Possible physical origin could be attributed to sensitivity of the dielectric response to particular type of ionic displacements.

## Acknowledgements

The generous help of E.Farhi with the instrument alignment is highly appreciated. S.V. thanks ILL for their hospitality and for the use of their facilities. The work is supported by the RFBR (grant 02-02-16695), CRDF (Project RP1-2361-ST-02) and the Russian Programme "Neutron Research of Solids".

**Single Column Figure/Scheme**

X

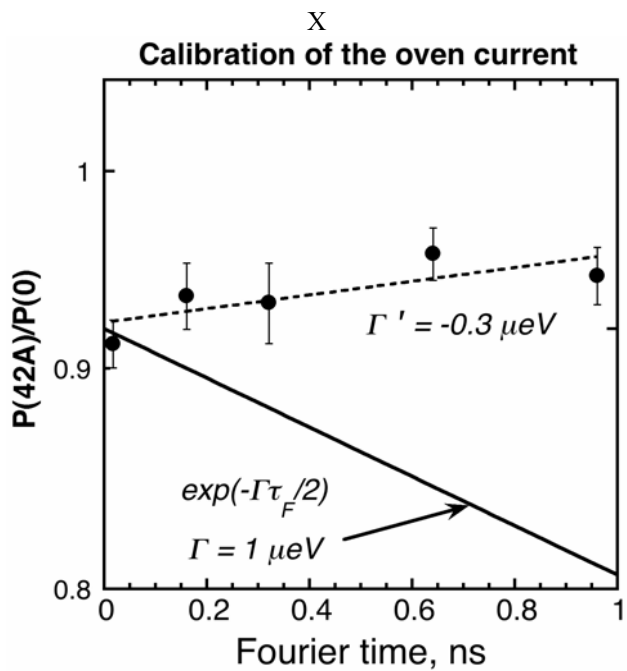

**Fig.1** Ratio of two spin-echo polarization contrasts (log scale) measured with heating current 42A and without current. The solid line for $\Gamma = 1$ μeV corresponds to the estimated instrument resolution.

X

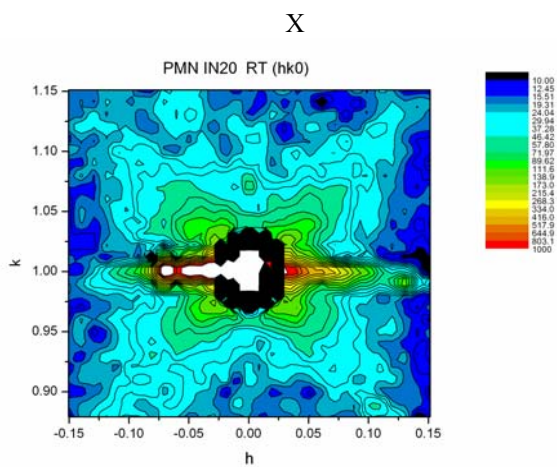

**Fig.2a** 2D scan of the diffuse scattering at room temperature in the vicinity of the Bragg point $\boldsymbol{\tau} = (0\ 1\ 0)$ measured with k = 2.662 Å$^{-1}$.

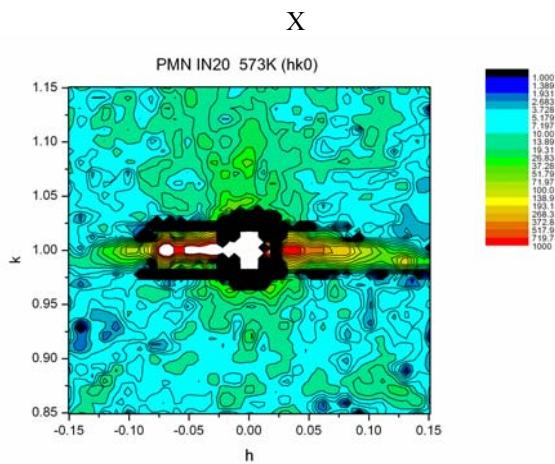

**Fig.2b** 2D scan in the vicinity of the Bragg point $\tau = (0\ 1\ 0)$ measured at T = 573 K with k = 2.662 Å$^{-1}$.

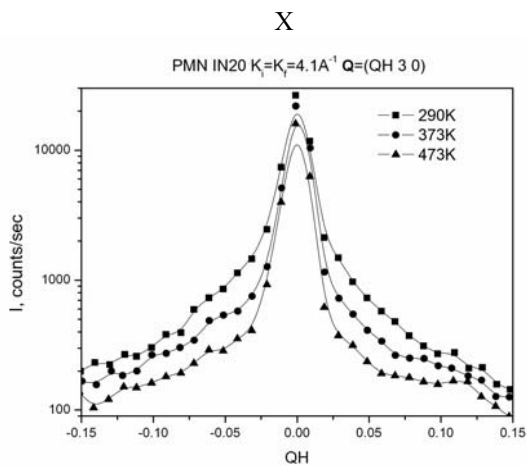

**Fig.3a** Temperature dependence of the scattering intensity measured with k = 4.1 Å$^{-1}$ along $Q_H$ in the vicinity of the Bragg point $\tau = (0\ 3\ 0)$. Note logarithmic scale on the intensity axis. Lines are guides to the eye.

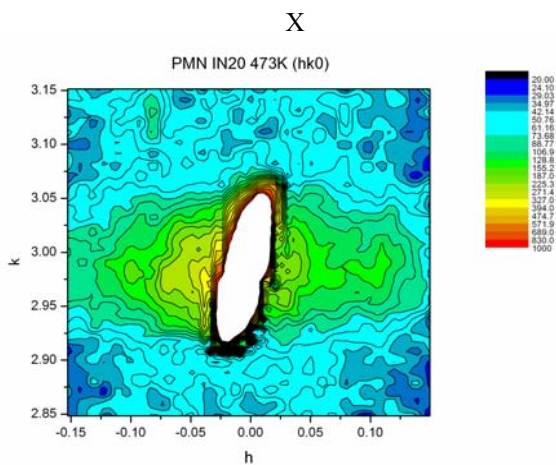

**Fig.3b** 2D scan of the scattering intensity in the vicinity of the Bragg point $\tau = (0\ 3\ 0)$ measured at T = 473 K with k = 4.1 Å$^{-1}$. The anisotropic diffuse scattering component is suppressed at this temperature while the transverse one is clearly visible.

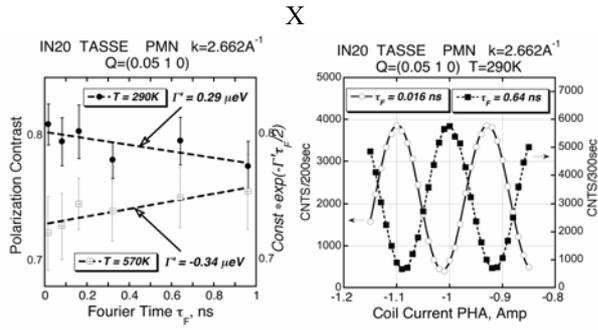

**Fig.4a** Left panel: Spin-echo polarization contrast (log scale) measured on scattering from crystal imperfections at room temperature and 573 K. No correction for the heating current is applied (to be compared with the fig.1). Fitted effective values of quasi-elastic widths are given. Right panel: Sinusoidal spin-echo signal at different Fourier times measured at 290 K as function of electric current in one of the compensating coils.

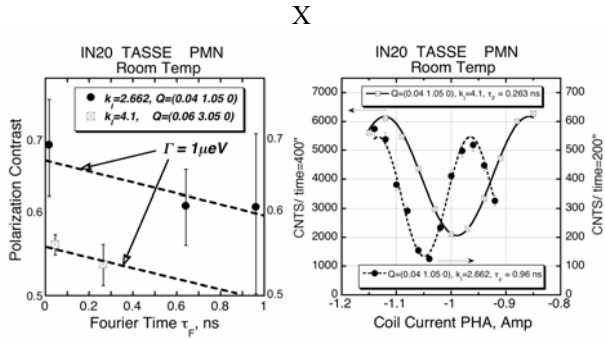

**Fig.4b** Left panel: Spin-echo polarization contrast (log scale) measured on scattering from butterfly-like component at room temperature with different $k_I$ and $Q$. Broken lines correspond to the 1 μeV quasi-elastic width. Difference in error bars is due to different statistics in the zones (010) and (030). Right panel: Sinusoidal spin-echo signal. Note 10–fold difference in the intensity scales.

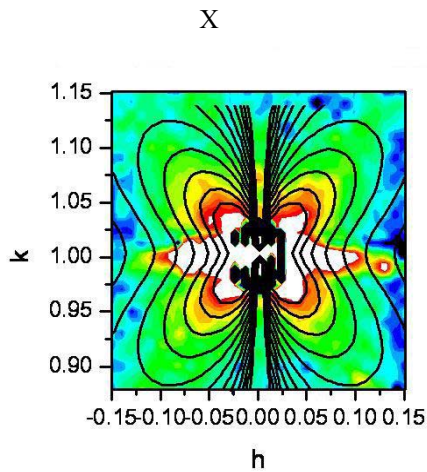

**Fig.5** Experimental intensity map (color) and calculated Huang scattering profile (lines). The butterfly component is well reproduced while the transverse scattering is absent in the calculated pattern.

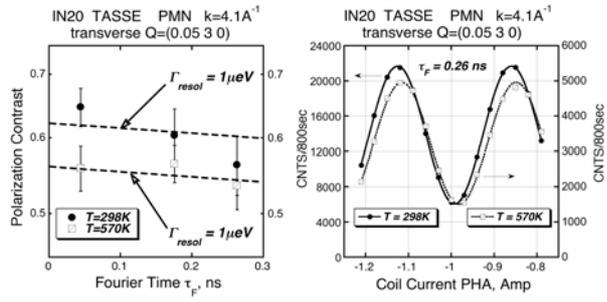

**Fig.6** Left panel: Spin-echo polarization contrast (log scale) measured with $k_I = 4.1$ Å$^{-1}$ on the transverse scattering component near Bragg point (030) at room temperature and 570 K. No correction for the heating current is applied (to be compared with the fig.1). Slope of the dashed lines corresponds to the resolution-defined 1 µeV quasi-elastic width. Right panel: Sinusoidal spin-echo signal at maximal Fourier time measured at room temperature and 570 K. Note the difference in the intensity scales.